# HIGHER ORDER MODES IN THIRD HARMONIC CAVITIES FOR XFEL/FLASH

I.R.R. Shinton[†*], N. Baboi[‡], N. Eddy[°], T. Flisgen[¥], H.W. Glock[¥], R.M. Jones[†‡], N. Juntong[†‡], T.N. Khabiboulline[°], U van Rienen[¥], P. Zhang[†‡*]

[†]School of Physics and Astronomy, The University of Manchester, Manchester, U.K.
[*]The Cockcroft Institute of Accelerator Science and Technology, Daresbury, U.K.
[‡]DESY, Hamburg, Germany.
[¥]Universität Rostock, Rostock, Germany.
[°]Fermilab, Batavia, USA.

## Abstract

We analyse higher order modes in the 3.9 GHz bunch shaping cavities recently installed in the FLASH facility at DESY. We report on recent experimental results on the frequency spectrum from probe based measurements made at CMTB at DESY. These are compared to those predicted by finite difference and finite element computer codes. This study is focused mainly on the dipole component of the multi-pole expansion of the wakefield.. The modes are readily identifiable as single-cavity modes provided the frequencies of these modes are below the cut-off of the inter-connecting beam pipes. The modes above cut-off are coupled to the 4 cavities and are distinct from single cavity modes.

## INTRODUCTION

The FLASH facility at DESY [1], and the European XFEL, currently under construction [2] utilises ultra-short bunches with a high peak current to generate coherent light with a concomitant high brilliance. These bunches are compressed by guiding them through a magnetic chicane and accelerating them in the TESLA cavities through an rf field below the crest or peak of the wave with a 1.3 GHz frequency. The non-flat field inevitably leads to tails in the bunch distribution. This has the detrimental effect of limiting the peak current. It is desirable to reduce these tails by flattening the overall field and this can be achieved by including harmonics of the fundamental frequency of the linac. A single frequency operating at the $n^{th}$ harmonic can be used to flatten out the dependence of the energy gain versus phase, by cancelling the second derivative of the fundamental at its peak. In practice at FLASH the first component in a Fourier expansion is used, namely the 3rd harmonic. A module consisting of four nine-cell cavities operating at 3.9 GHz has been designed and built by FNAL (illustrated in Fig. 1) and has recently been tested and commissioned at DESY [3]. Some essential parameters for this cryo-module are displayed in Table 1. The wakefields in these third harmonic cavities are considerably larger than those in the main accelerating linacs as the iris radius is significantly smaller (15 mm compared to 35 mm in TESLA). It is important to damp the modal components of the wakefields and to ensure there are no trapped modes with particularly high R/Q values. An experimental and simulation study is underway to characterise these modes and to instrument the cavity with electronics suitable for beam-based diagnostics. In this manner, the modes will also be used to remotely align the beam to the electrical centre of the cavity and to ascertain internal misalignments.

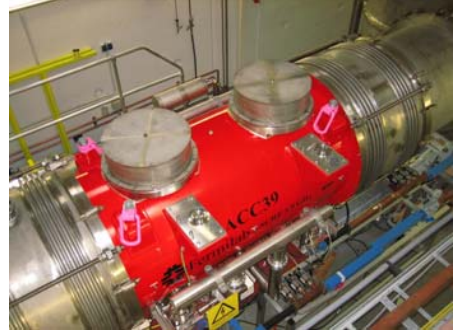

Figure 1: FNAL ACC39 Cryo-module [4] consisting of four 3.9 GHz cavities.

| Number of cavities | 4 |
|---|---|
| Cavity length | 0.346 m |
| Design gradient | 14 MV/m |
| Phase | -179° |
| Cavity frequency ($\omega/2\pi$) | 3.9 GHz |
| R/Q ($V^2/(2\omega U)$) [8] | 373 Ω |
| $E_{peak}/E_{acc}$ | 2.26 |
| $B_{peak}$ ($E_{acc}$=14MV/m) | 68 mT |
| Loaded $Q_L$ | $1.3 \times 10^6$ |
| Max Dipole R/Q at $\omega/2\pi$ | 50.20 Ω/cm², 4.831 GHz |

Table 1: Cryo-module ACC39 parameters

Finally we note that flattening the field also reduces the growth of transverse phase space. The transverse magnetic fields arise from the rate of change of the longitudinal electric field. Thus, flattening the electric field will also result in a reduced magnetic field. Hence the use of a cryo-module of third harmonic cavities will reduce the dilution of both longitudinal and transverse phase space.

Prior to developing the electronics for beam diagnostics, it is important to characterise the modes within the cavities. For this purpose we investigated the spectra of these modes, measured within the Cryo-Module Test Bench (CMTB) facility at DESY. Monopole, dipole, quadrupole and sextupole modes have been studied. We report on measurements made at FLASH in the next section. In addition, after installation in FLASH, we also made measurements on the modes and compared them to simulations. As the inter-cavity connecting beam tubes are above cut-off, the modes from each cavity are coupled and are quite distinct from individual cavity modes. These effects are also studied. This paper concludes with some final remarks on the present study and on additional work planned.

## MEASUREMENTS OF HOMS AT FLASH

After mode characterisation in CMTB, the cryo-module was commissioned and installed [3] at the FLASH facility in DESY. ACC39 was then supplied with HOM cables and the spectra were obtained in a similar manner as was applied in CMTB. In recording the spectra, monopole, dipole and quadrupole modes were identified. We also encountered modes arising due to the coupling of inter-connected cavities, as well as the usual discrete cavity modes. A description of the measurement and procedure followed is now provided.

ACC39 is composed of four interconnected cavities, referred to as C1 through C4. There are two distinct coupler designs: a 1-leg-coupler design (C1 and C3) and a 2-leg-coupler design (C2 and C4), this is illustrated in Fig. 2. Measurements across the string of cavities within ACC39 were conducted using a Rohde and Schwarz-ZVA8 vector network analyser taken at the ACC39 HOM board rack. Both transmission and reflection data were recorded from the upstream and downstream HOM ports in the beam direction as is indicated in Fig. 2, such that the complete scattering matrix of the cavity string was sampled moving from C1 to C4 (resulting in 36 sets of scattering data, including measurements across the beam-pipes).

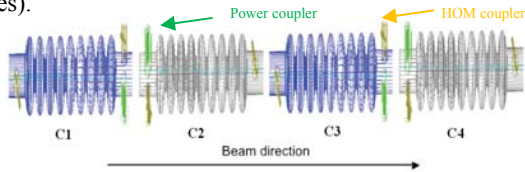

Figure 2: Schematic representation of 4 cavities within ACC39 [5]. The power couplers are indicated in green. C1 and C3 contain the 1-leg-HOM couplers whilst C2 and C4 contain the 2-leg-HOM couplers.

In the probe-based transmission and reflection measurements we scanned over a frequency range of 3.5 GHz to 8 GHz with a step of 10 kHz. The long cables connecting the HOM ports of ACC39 to the instrumentation racks inevitably added ~ 10 dB of losses and phase delays to the cavity measurements. Calibration measurements for the cables have been performed, but due to time constraints are not reported here. We recorded the spectra for all cavities. A typical spectrum on the transmission properties of cavity 1, together with simulations indicating the single-cavity band structure, is illustrated in Fig. 3. Monopole, dipole and quadrupole modes are present in this 8 GHz scan.

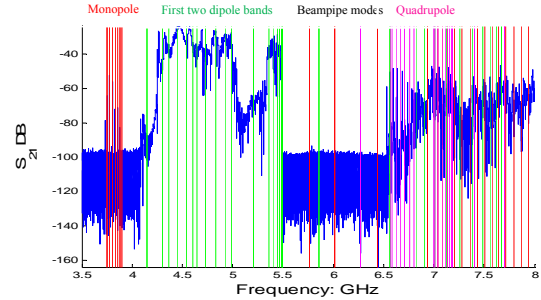

Figure 3: Transmission data taken across C1 (from the downstream HOM of C1 to the upstream HOM of C1). The vertical lines indicate the simulations performed with MAFIA [6]. The colours red, green magenta indicate monopole, dipole and quadrupole modes, respectively.

Detailed investigations on the spectra indicate that the modes above the cut-off frequency of the beam tubes (~ 4.39 GHz for dipole modes) are shifted in frequency with respect to their isolated cavity values. This is a possible indication of multi-cavity modes being present in the module; although fabrication tolerances could well contribute to these shifts. An example of such a multi-cavity dipole mode is displayed in Fig. 4. Quadrupole modes are also readily identifiable in this structure and they bear comparison to simulations of single cavity simulations. The majority of these quadrupole modes are within ~ 15 MHz of their isolated cavity counterparts.

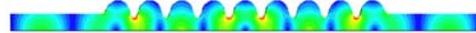

Figure 4: Simulation, made with HFSS [7], of the first dipole band of the electric field at 4.831 GHz with an R/Q 50.20 Ω/cm2 [8].

Additional measurements were conducted on the structure and compared to simulations. These are reported on in the next section.

## MEASUREMENTS AND SIMULATION OF MODE SPECTRA

The mode spectra of several cavities were characterised. A selected result, again based on un-calibrated measurements, is displayed in Fig. 5, together with a prediction based on simulations made with CST Microwave Studio [9], analytical calculations and cascading simulations. The *Fast Resonant* solver module of Microwave Studio was used in the simulation of the main accelerating cavities. Couplers and bellows were also simulated. Waveguides, azimuthal rotation of modes and cables connecting the ports were taken into account analytically. All these individual components are then cascaded together, retaining 8 propagating modes, to enable the overall scattering matrix to be obtained [10], [11]. Considering the number of quantities which are not known accurately, such as the effect of tuning the cavities on the cavity geometry and the assumption of a perfectly matched power couplers, the degree which the prediction agrees with the experimental result is quite remarkable. Additional work is in progress to take into account some of these additional effects on the mode spectra.

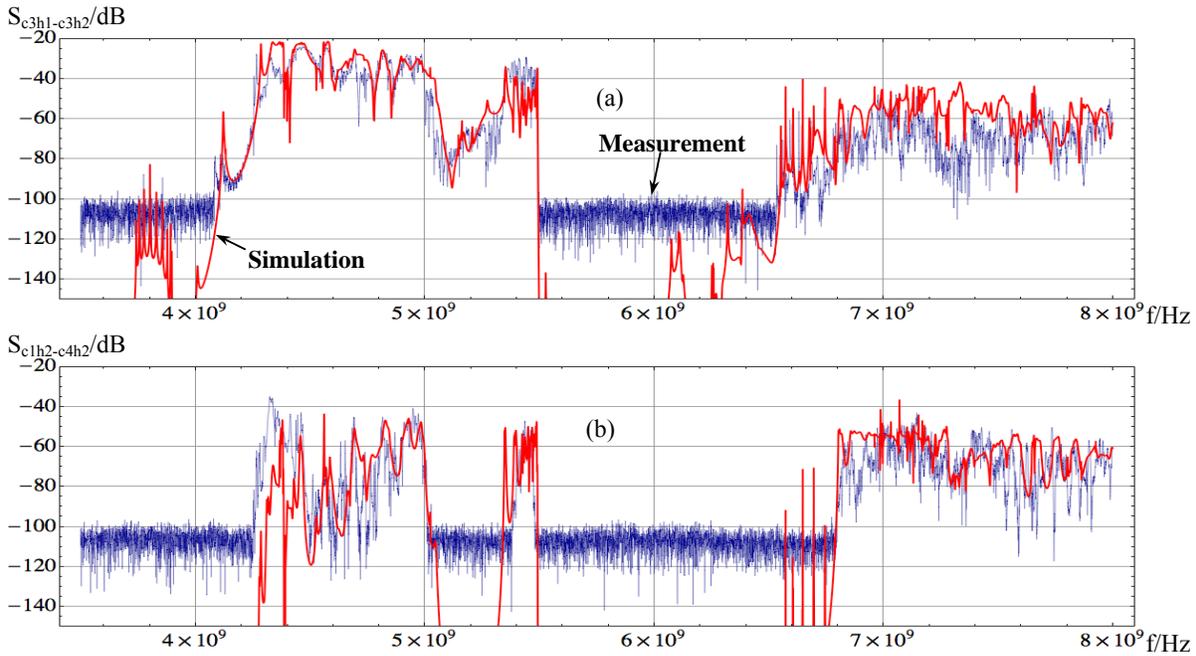

Figure 5: Comparison of $S_{21}$ measurements (blue) and simulations (red). Transmission through C3, upstream to downstream HOM-coupler (a). Transmission from C1, upstream HOM-coupler, to C4, downstream HOM-coupler (b).

## FINAL REMARKS

Third harmonic cavities have been installed at FLASH in order to linearise the rf field and hence improve the peak current of the XFEL facility at DESY. The goal of our work is to instrument the cavities with beam diagnostics. To this end, we have made an initial investigation into the modal spectra of the cavities. Prior to and after installation in FLASH, the mode spectra have been obtained in CMTB and in FLASH without beam excitation. The frequencies of modes above the cut-off of inter-connecting beam tubes are shifted with respected to their single cavity values due to mode coupling. Initial simulations on the dipole mode spectra are in reasonable agreement with combined finite element and CSC simulations. Future work will be concerned with an additional characterisation of the dipole modes both with and without beam-excitation, with a view to determining a mode suitable for the electronics for beam and cavity alignment studies.

## ACKNOWLEDGEMENTS

This research has received funding from the European Commission under the FP7 Research Infrastructures grant no. 227579. N.J. receives support from the Royal Thai Government and the Thai Synchrotron Light Research Institute. We are pleased to acknowledge H. Ewald, for providing the R&S-ZVA8-network analyser and D. Mitchell, W.D. Möller and E. Vogel [12] for contributing important geometrical details on the cavities and coupler within ACC39.

## REFERENCES

[1] S. Schreiber et al., EPAC08, MOPC030
[2] M. Altarelli et al. (eds), DESY-06-097, 2006
[3] E. Vogel et al., IPAC 2010, THPD003
[4] T. Khabibouline et al., TESLA-FEL-2003-01, 2003.
[5] T. Khabiboulline et al., Private communication
[6] MAFIA©, Ver. 4.106, CST AG, Darmstadt, Germany
[7] HFSS, Ver.11 2090, Ansoft, USA
[8] I.R.R.Shinton et al, Mode distribution in the third harmonic FLASH/XFEL cavities, CI Int.Note
[9] CSTStudio©, Ver. 2010, CST AG, Darmstadt, Germany
[10] K. Rothemund et al., TESLA-REPORT 2000-33
[11] H.W. Glock et al., TESLA-REPORT 2001-25,
[12] E. Vogel et al. Private communication.